\documentclass[sn-mathphys,Numbered, iicol]{sn-jnl}

\usepackage{graphicx}%
\usepackage{multirow}%
\usepackage{amsmath,amssymb,amsfonts}%
\usepackage{amsthm}%
\usepackage{mathrsfs}%
\usepackage[title]{appendix}%
\usepackage{xcolor}%
\usepackage{textcomp}%
\usepackage{manyfoot}%
\usepackage{booktabs}%
\usepackage{algorithm}%
\usepackage{algorithmicx}%
\usepackage{algpseudocode}%
\usepackage{listings}%

\usepackage{subcaption}
\usepackage{multicol}

\usepackage{xcolor}%

\begin{document}

\title[Article Title]{Online Clustering of Known and Emerging Malware Families}

\author*[1]{\fnm{Olha} \sur{Jure\v{c}kov\'{a}}}\email{jurecolh@fit.cvut.cz}
\author[1]{\fnm{Martin} \sur{Jure\v{c}ek}}\email{martin.jurecek@fit.cvut.cz}

\author[2]{\fnm{Mark} \sur{Stamp}}\email{mark.stamp@sjsu.edu}

\affil*[1]{\orgdiv{Faculty of Information Technology}, \orgname{Czech Technical University in Prague},\\ \orgaddress{\city{Prague}, \country{Czechia}}}
\affil[2]{\orgdiv{Department of Computer Science}, \orgname{San Jose State University}, \orgaddress{\city{San Jose}, \state{California}, \country{USA}}}

\abstract{Malware attacks have become significantly more frequent and sophisticated in recent years. Therefore, malware detection and classification are critical components of information security. Due to the large amount of malware samples available, it is essential to categorize malware samples according to their malicious characteristics. Clustering algorithms are thus becoming more widely used in computer security to analyze the behavior of malware variants and discover new malware families. Online clustering algorithms help us to understand malware behavior and produce a quicker response to new threats. This paper introduces a novel machine learning-based model for the online clustering of malicious samples into malware families. Streaming data is divided according to the clustering decision rule into samples from known and new emerging malware families. The streaming data is classified using the weighted $k$-nearest neighbor classifier into known families, and the online $k$-means algorithm clusters the remaining streaming data and achieves a purity of clusters from 90.20\% for four clusters to 93.34\% for ten clusters. This work is based on static analysis of portable executable files for the Windows operating system. Experimental results indicate that the proposed online clustering model can create high-purity clusters corresponding to malware families. This allows malware analysts to receive similar malware samples, speeding up their analysis.}


\maketitle

\section{Introduction}

In the field of malware detection, there are usually two sides. One party participates in malware creation, while its primary purpose is profit \cite{huang2018systematically}. The other side detects the malware and tries to minimize the damage. In the past, malicious programs were written by hand, which was time-consuming. In addition, in-depth knowledge of operating systems, networks, programming, and others, was required to create the malware. Today, the creation of malicious programs is fast, and it is not even necessary to have the mentioned theoretical knowledge. There are several programs that facilitate the creation of malware. These are malware generators, defined as programs that receive a set of parameters $S$ as an input, and malware will be generated as an output. Some of these programs are also freely available, mainly for scientific purposes. However, most of them are difficult to access, e.g., located on the darknet, and some money may be required to provide the malware generator.

For a given malware generator $G$ and a set $S$ of specific parameters $\{p_1, \ldots, p_n\}$, it is then possible to generate particular malware $m_1$. For a different set of parameters $\{q_1, \ldots, q_n\}$, the same generator $G$ generates a different malware $m_2$. Depending on the particular generator $G$, it is possible to specify the differences between $m_1$ and $m_2$ based only on the parameter sets. For example, programs $m_1$ and $m_2$ can perform the same harmful activity and differ in obfuscation techniques. Both malware can perform various harmful activities, e.g., one malware may be aimed at stealing passwords and the other at blocking access to the system, both generated from the same generator. For this reason, for the sake of simplicity, we will generally consider malware generators as programs that generate some malware for a given set of parameters.

Based on various analyses performed on real malware generators, researchers hypothesized that malware samples generated from the same generator are similar \cite{wong2006hunting, nataraj2011malware, cesare2011malware, jurecek2021improving}. More precisely, for an appropriate distance, elements generated by the same generator with a given set of parameters $S$ are close to each other. In this case, these elements come from the same malware family. So some generators, with their parameter sets, can be identified with malware families. Many papers \cite{rudd2024efficient, chen2022malware} on malware classification are built on the assumption that malware samples from one family are close to each other and distant from different malware families and benign files. Solving the problem of classifying malware into malware families has practical applications in antivirus companies. These companies receive hundreds of thousands of new malicious samples daily \cite{avtest2024avtest}, which are either processed manually by malware analysts or automatically using detection systems usually based on machine learning. Suppose it was possible to group malware samples into groups based on appropriate similarity. In that case, it is possible that elements from the same groups would essentially belong to the same malware family. Thus, malware analysts could receive similar malware samples, speeding up their analysis.

Malware clustering is also necessary for scientific purposes since it provides the knowledge necessary for the examination of the evolution of individual malware families over time. This research might then be used to predict future variants of malware. This area is important for the antivirus industry since it can help reduce the so-called reaction time, defined as the period between spreading the malware by some infection vector, finding the malware, and creating a detection rule for it.

This paper presents a new machine learning-based model for the online clustering of malicious samples into malware families. Our proposed online clustering algorithm can cluster samples one by one based on already clustered samples and does not need to have all samples available immediately. We designed a new clustering decision rule to determine which incoming samples belong to known or new emerging malware families. These two groups are then processed online, and our experimental results show that this approach is more successful in terms of the purity of clusters than the approach where we directly apply the online clustering algorithm.

This paper is organized as follows. Section \ref{related_work} reviews related works on malware family clustering, and Section \ref{background} presents three online clustering algorithms used in the experimental part. Section \ref{proposed} presents the proposed online clustering system and our experimental setup. Section \ref{evaluation} describes the experimental results. Finally, Section \ref{conclusion} concludes the paper and presents suggestions for future work.

\section{Related Work} \label{related_work}

There is a growing interest in the use of unsupervised methods in malware detection, image processing, and wireless communication, for example. This section presents recent works that dealt with malware detection or classification using unsupervised learning methods.
 
In \cite{pitolli2021malfamaware}, the authors propose MalFamAware, an online clustering method for incremental automatic malware family identification and malware classification. This method effectively updates the clusters when new samples are added without having to rescan the entire dataset. The authors use BIRCH (Balanced Iterative Reducing and Clustering using Hierarchies) as an online clustering algorithm. It is compared with CURE (Clustering using Representatives), DBSCAN, $k$-means, and other clustering algorithms. MalFamAware either classifies new incoming malware into the corresponding existing family or creates a class for a new family, depending on the situation.

The authors of \cite{soufiane2019clustering} propose a clustering method based on incremental learning. This method is based on two-phase clustering. A clustering ensemble method is used to group the dataset objects to complete the first phase. The final clustering result is then extracted using an incremental clustering algorithm in the second phase. The authors use three clustering algorithms: $k$-means, partitioning around medoid, and self-organizing maps (SOM) with different random initializations and the voting mechanism to extract a set of sub-clusters.

The authors of \cite{hua2020clustering} propose the clustering ensemble method, an extension of the self-organizing map combined with the cascaded structure, also known as a cascaded SOM. The method cascades the outputs of multiple SOM networks and uses them as input to another SOM network.

A framework based on an unsupervised machine learning algorithm called "SOMDROID" is proposed by the authors of \cite{mahindru2022somdroid}. They use SOM to create a model to determine whether an Android app is benign or malicious. The authors use six different feature ranking approaches to select significant features or feature sets and then apply the self-organizing map algorithm to the selected features or feature sets.

In \cite{zhuang2012ensemble}, the authors create an automatic categorization system to automatically group phishing websites or malware samples into families with common characteristics using a cluster ensemble. Their approach combines the individual clustering solutions produced by different algorithms using a cluster ensemble. The authors use the $k$-medoids and hierarchical clustering algorithms to create the base clusterings. 

The authors of \cite{wilkins2020cougar} create and test a new system called COUGAR (Clustering of Unknown malware using Genetic Algorithm Routines), which uses a multi-objective genetic algorithm to reduce high-dimensional malware behavioral data and optimize clustering behavior. The EMBER (Endgame Malware Benchmark for Research) dataset is used, and the dimensionality reduction method chosen for it is UMAP (Uniform Manifold Approximation and Projection). Although this method can be parameterized to reduce to any number of dimensions, the two-dimensional embedding reduction is selected for this paper due to its simplicity and ease of visualization. The authors use three clustering algorithms: DBSCAN, OPTICS, and $k-$means and Non-dominated Sorting Genetic Algorithm III (NSGA-III). The optimal parameters for each clustering algorithm are determined by training them on 2,000 samples from EMBER. This procedure is repeated ten times in order to account for the stochastic nature of genetic algorithms. The authors also investigate a hypothetical situation by applying the system to a realistic, real-world scenario.

The authors of \cite{pai2017clustering} investigate the problem of malware classification using the $k$-means and Expectation-Maximization (EM) clustering algorithms. They use Hidden Markov Models (HMM) to generate the scores for the clustering techniques. The authors create clusters from HMM scores using both $k$-means and the EM clustering algorithm. The authors use the silhouette coefficient to evaluate the clustering results. In addition, they use a simple purity-based score to determine clustering success. In their research, the authors focus primarily on the three dominant families in the Malicia dataset: Zbot, ZeroAccess, and Winwebsec.

In \cite{basole2021cluster}, the authors examine the relationship between malware families. The features they employ for clustering are based on byte n-gram frequencies, and they use the $k-$means algorithm as their clustering technique. The authors analyze a dataset that contained 1,000 samples from 20 malware families, which can be categorized into seven different malware types. They present three distinct sets of clustering experiment results. The authors first cluster every pair of malware families, then investigate clustering experiments in which they focus on a single family of each malware type under consideration, and finally consider clustering multiple families from the same malware type. The authors use the adjusted Rand index (ARI) to evaluate the clustering results.

The authors of \cite{pirscoveanu2016clustering} use SOM to generate clusters that capture similarities between malware behaviors. Pirscoveanu et al. use features chosen based on API calls, which represent successful and unsuccessful calls (i.e., calls that have succeeded, resp. failed in changing the state of the system on the infected machine) and the return codes from failed calls. The authors use principal component analysis to reduce the set of features and the elbow method and gap statistics to determine the number of clusters. Each sample is then projected onto a two-dimensional map using self-organizing maps, where the number of clusters equals the number of map nodes. The dataset is used to generate a behavioral profile of the malicious types, which is then passed to a self-organizing map, which compares the proposed clustering result with labels obtained from Antivirus companies via VirusTotal\footnote{\url{https://www.virustotal.com}}. 

The relay selection algorithm and the power control protocol presented by the authors of \cite{hajjar2019relay} are based on the Basic Sequential Algorithmic Scheme (BSAS) and do not require any additional infrastructure, in contrast to other capacity-improving techniques. Users will instead act as temporary relay stations. The authors modify the original BSAS to fit the requirements of power control and resource allocation while also making it suitable for an LTE environment. The newly proposed BSAS-based algorithm uses path loss as the proximity between a node and formed clusters instead of using distance. The fundamental concept is that, based on its path-loss from the previously formed clusters, each node is assigned to either one that already exists or one that has just been formed.

The authors introduce a hybrid model of AE and SOM to detect IoT malware in \cite{nguyen2022denoising}. The proposed models are evaluated using the NBaIoT dataset in various aspects, including detecting new or unknown malware, transferring knowledge for detecting IoT malware on various IoT devices, and detecting different IoT malware groups. The authors also examine the latent representation of DAEs (Denoising AutoEncoder) for unsupervised learning in IoT malware detection. There are two stages to the newly proposed hybrid model for identifying IoT malware. To create its latent representation, DAE is trained on unlabeled data in the first phase, which includes both normal data and IoT malware. During the second phase, the SOM functions as a method for classification that works directly with the feature space of the DAE.

\section{Theoretical Background} \label{background}

Clustering algorithms are unsupervised machine learning methods that aim at grouping abstract objects into clusters of similar objects. This work focuses on online clustering algorithms, which are computational procedures that process streaming data incrementally as data points arrive over time. This section presents three state-of-the-art online clustering algorithms used in the experimental part: Online $k$-means (OKM), Basic Sequential Algorithmic Scheme (BSAS), and Self-Organizing Map (SOM). We applied all these algorithms to cluster the samples into malware families. At the end of this section, the distance-weighted $k$-nearest neighbor classifier, which is included in our proposed model, is briefly presented.

\subsection{Online $k$-means (OKM)}
The online $k$-means (OKM) algorithm, also known as sequential $k$-means or MacQueen's $k$-means \cite{abernathy2022incremental} is an example of a non-hierarchical clustering algorithm. The sequential $k$-means algorithm sequentially clusters a new example and updates a single center immediately after a data point is assigned to it. The number of clusters, $k,$ must be determined in advance, which is one disadvantage of the online $k$-means algorithm. The pseudocode for the online $k$-means algorithm is given in Algorithm \ref{seqkmeans} below \cite{duda_skmeans}.

\begin{algorithm}
\caption{Sequential  $k$-means algorithm (OKM)}
\label{seqkmeans}
\begin{algorithmic}[1]
\Require a number of clusters $k$ to be created, a set of data points $X$
\Ensure a set of $k$ clusters
\State initialize cluster centroids $\mu_1, \ldots, \mu_k $ randomly 
\State set the counts $n_1, \ldots, n_k$ to zero 
\Repeat
\State select a random point $x$ from $X$ and find the

\hspace{-0.25cm} nearest center $\mu_i$ to this point
\If{$\mu_i$ is closest to $x$}
\State increment $n_i$
\State replace $\mu_i$ by $\mu_i + \frac{1}{n_i}( x - \mu_i)$
\EndIf
\Until{interrupted} %
\end{algorithmic}
\end{algorithm}

\subsection{Self-organizing Map (SOM)}
In 1982, Teuvo Kohonen introduced the concept of self-organizing maps, or SOMs. Consequently, they are occasionally referred to as Kohonen maps \cite{kohonen1990self}. The SOM is an unsupervised machine learning technique that preserves similarity relations between the presented data while converting a complex high-dimensional input space into a simpler low-dimensional (typically two-dimensional grid) discrete output space. Self-organizing maps use competitive learning rules in which output neurons fight with one another to be active neurons, activating only one of them at a time. A winning neuron is an output neuron that has won the competition.

Before running the algorithm, several parameters need to be set, including the size and shape of the map, as well as the distance at which neurons are compared for similarity. After selecting the parameters, a map with a predetermined size is created. Individual neurons in the network can be combined into layers. 

SOM typically consists of two layers of neurons without any hidden layers \cite{asan2012introduction}. The input layer represents input vector data. A weight is a connection that connects an input neuron to an output neuron, and each output neuron has a weight vector associated with it. The formation of self-organizing maps begins by initializing the synaptic weights of the network. The weights are updated during the learning process. The winner is the neuron whose weight vector is most similar to the input vector.

The winning neuron of the competition or the best-matching neuron $c$ at iteration $t$ (i.e., for the input data $x_t$) is determined using 
$$c(t)=\arg \min\left\{ \left\| x(t) - w_{i}(t)\right\|\right\}, \textrm{ for } i=1,2,\ldots,n$$
where $w_{i}(t)$ is the weight of $i$-th output neuron at time $t$, and $n$ is the number of output neurons. After the winning neuron $c$ has been selected, the weight vectors of the winner and its neighboring units in the output space are updated. The weight update function is defined as
$$w_i(t+1)=w_i(t)+\alpha(t)h_{ci}(t)\left[x(t)-w_i(t) \right],$$
where $\alpha(t)$ is the learning rate parameter, and $h_{ci}(t)$ is the neighborhood kernel function around the winner $c$ at time $t.$ The learning rate is the speed with which the weights change.  
The connection between the input space and the output space is created by the neighborhood function, which also determines the rate of change of the neighborhood around the winner neuron. This function affects the training result of the SOM procedure. 

A Gaussian function is a common choice for a neighborhood function
$$h_{ci}(t)=\exp \left(-\frac{d_{ci}^2}{2{\sigma}^2(t)} \right)\alpha(t).$$

\noindent that determines how a neuron is involved in the training process, where $d_{ci}$ denotes the distance between the winning neuron $c$ and the excited neuron $i$, ${\sigma}^2(t)$ is a factor used to control the width of the neighborhood kernel at time $t.$ The learning rate $\alpha(t)$ is a decreasing function toward zero.

There are many applications for SOM, and one of them is clustering tasks. The authors of \cite{baccao2005self} claim that since each SOM unit is the center of a cluster, the $k$-unit SOM successfully finished a task comparable to $k$-means. The authors further stated that the SOM and $k$-means algorithms strictly correspond to one another when the radius of the neighborhood function in the SOM is zero.

The basic SOM algorithm can be summarized by the following pseudocode:
\begin{algorithm}
\caption{Self-organizing map (SOM)}
\label{som}
\begin{algorithmic}[1]
\Require dimension and size of the output space, distance function, neighborhood function, learning rate, and a set of data points $X$.
\Ensure a set of clusters
\State initialize the weights of each neuron 
\State $t=1$ 
\State select randomly an input vector from the set of training data $X$
\For {each input vector} 
\State calculate the distances measure between  

\hspace{-0.25cm} the input vector and all the weights 

\hspace{-0.25cm} vectors. 
\State find the best matching neuron $c(t)$ at 

\hspace{-0.25cm} iteration $t$.
\State update the weight vectors of the neurons.
\State $t=t+1$ and update neighborhood size and

\hspace{-0.25cm}  learning rate.
\EndFor  %
\end{algorithmic}
\end{algorithm}

\subsection{Basic Sequential Algorithmic Scheme (BSAS)}
The following algorithm we employed in our work is the Basis Sequential Algorithmic Scheme (BSAS), a sequential clustering technique that presents all feature vectors to the algorithm once \cite{koutroumbas2008pattern}. The number of clusters is unknown in advance. Clusters are gradually generated as the algorithm evolves. The basic idea behind BSAS is to assign each newly considered feature vector $x$ to an existing cluster or to create a new cluster for that vector based on the distance to previously created clusters. To determine whether a data point can join a particular cluster, the algorithm considers two thresholds: a maximum number of clusters that can be merged and a dissimilarity threshold.

There are several ways to define the distance $d(x,C)$ between a cluster $C$ and a feature vector $x$. We will consider $d(x,C)$ as the distance between $x$ and the centroid of $C$. The parameters of the BSAS are as follows: a number $q$, which represents the maximum number of clusters permitted, and a dissimilarity threshold $\Theta$, which is the threshold used for creating new clusters. A new cluster with the newly presented vector is formed when the distance between a new vector and any other clusters is beyond a dissimilarity threshold and if the number of the maximum clusters allowed has not been reached. The threshold $\Theta$ directly affects the number of clusters formed by BSAS. If the user chooses the too small value of  $\Theta,$ then unnecessary clusters will be created, while if the user chooses the too large value of $\Theta,$ less than an appropriate number of clusters will be formed. The pseudocode for the BSAS algorithm is given below in Algorithm \ref{bsas}.

\begin{algorithm}
\caption{Basic Sequential Algorithmic Scheme (BSAS)}
\label{bsas}
\begin{algorithmic}[1]
\Require the dissimilarity threshold $\Theta$, the maximum allowed number of clusters $q$, and a set of data points $X$
\Ensure a set of clusters
\State initialize $m=1$ 
\State select a random point $x_1$ from $X$ 
\State define the first cluster $C_m=\left\{ x_1 \right\}$
\For {\textbf{each} $x$  \textbf{in}  $X$\textbackslash$\{x_1\}$} 
\State find $C_k: d(x,C_k)=min_{1\leq i \leq m} d(x,C_i)$
\If{$d(x,C_k)>\Theta$ and $m<q$}
\State $m=m + 1$
\State $C_m= \left\{ x \right\}$
\Else
\State $C_k= C_k \cup \left\{ x \right\}$
\State update the centroid of $C_k$ 
\EndIf
\EndFor
\end{algorithmic}
\end{algorithm}

\subsection{Distance-weighted $k$-nearest Neighbor (WKNN)}

The distance-weighted $k$-nearest neighbor classifier (WKNN) \cite{dudani1976distance} is used in our work to classify testing data to known malware families. The main idea behind the WKNN is that closer neighbors have larger weights than neighbors far away from the query object. Let $T = (z_1, \ldots, z_k)$ be $k$ nearest neighbors from $D$ of the query object $x \in S$ and $d_1, \ldots, d_k$ the corresponding distances arranged in increasing order. The resulting cluster $C_{x}$ for $x$ is defined by the majority weighted vote
\begin{equation} \label{wknn}
C_{x} = \operatorname*{argmax}_C \sum_{(z_i,C_{z_i}) \in T} w_i \cdot \delta(C,C_{z_i}) 
\end{equation} 

\noindent where $(z_i,C_{z_i})$ denotes that the sample $z_i$ belongs to the cluster $C_{z_i}$, $\delta(a,b)$ is equal to one if \mbox{$a=b$} and zero otherwise, and the weight $w_i$ for $i$-th nearest neighbor $z_i$ is defined by

\begin{displaymath}
	w_i = \left\{ \begin{array}{ll}
\frac{d_k-d_i}{d_k-d_1} & \textrm{if $d_k \neq d_1$}\\
1 & \textrm{otherwise.}
\end{array} \right.
\end{displaymath}

\section{Proposed Approach and Experimental Setup} \label{proposed}

This section presents the proposed model for the online clustering of malicious samples to malware families and the experimental setup, which contains detailed information about the methodology and procedures used in experiments.

Suppose we have a dataset $D = \{x_1,\ldots ,x_t\}$, which contains $t$ unlabeled feature vectors for malware samples. Let $K$ be a set of all malware families that the samples from $D$ belong to. Suppose that we have chronologically ordered streaming data $S = \{x_{t+1}, x_{t+2}, \ldots\}$, which contains malware samples that belong to the set $K$ of malware families and newly emerging families. The goal is to cluster the data set $D$ together with the data set $S$ so that the clusters in each of these data sets achieve the highest possible purity and thus correspond closely to the malware families.

\begin{figure*}
\centering
\includegraphics[scale=0.9]{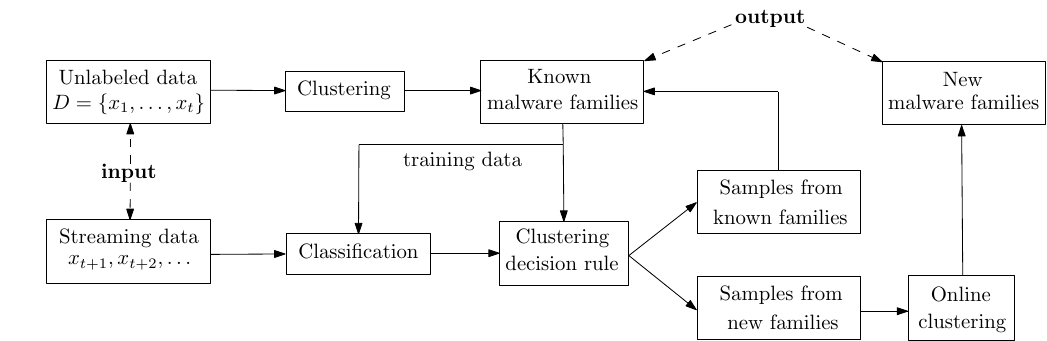}
\caption{The architecture of the proposed model for the online clustering of malicious samples to malware families.}
\label{fig:1}
\end{figure*}

This goal aims to simplify the work of malware analysts since they would receive samples from the same malware family, which would speed up the overall analysis process. The missing labels of samples from dataset $D$ corresponds to a real situation when antivirus companies received the newest samples, which had not yet been analyzed, i.e., they were not subjected to machine learning algorithms that could predict the labels, nor were they manually analyzed by malware analysts.

Therefore, we assume that when deploying the model proposed in this section,  we will have several unlabeled samples available. These samples can be clustered using a batch approach, where the clustering algorithm has all the samples available. We assume that all samples from dataset $D$ appeared before a specific time $T$. On the other hand, from the time $T$, new malware samples $S = {x_{t+1}, x_{t+2}, \ldots}$ arrive as streaming data. Since we will always have samples until the current time and we have to wait for newer samples, we use online clustering to cluster incoming samples from $S$. This type of clustering algorithm can cluster samples one by one based on already clustered samples and does not need to have all samples available immediately. Streaming data in the real world contain benign and malicious samples. However, in this work, we only work with malware samples, assuming that the benign samples of the streaming data $S$ have been filtered out.

\subsection{Proposed Model}

The proposed model for clustering samples from a fixed dataset $D$ and a streaming data $S$ is illustrated in Fig. \ref{fig:1}. Dataset $D$ is first preprocessed using the standard score and principal component analysis (PCA). The preprocessed dataset $D$ is then clustered using a clustering algorithm. In this work, we experimented with three clustering algorithms, and based on the results from Section \ref{clust}, we used the SOM algorithm. The samples from dataset $D$ are clustered into malware families from the set $K$, referred to as \textit{known malware families}.

The streaming data $S = {x_{t+1}, x_{t+2}, \ldots}$ is one by one preprocessed via the standard score, and PCA, using the same setup used for processing of dataset $D$. Then the incoming samples $x_{t+1}, x_{t+2}, \ldots$ will be clustered one by one according to the following approach. The sample $x \in S$ is first classified to the cluster $C_{x}$ from the clustering of dataset $D$ according to the WKNN classifier. Cluster names are used as labels for samples from dataset $D$, which is used to train the WKNN classifier. 

After the identification of $C_{x}$ for $x \in S$ using WKNN classification, the \textit{Clustering decision rule} determines whether $x$ will remain in the cluster $C_{x}$ of samples from \textit{known malware families} or will be assigned to some cluster of samples from \textit{new malware families}, i.e., families that appeared after the time $T$. The \textit{Clustering decision rule} is defined by \\

\noindent\textit{$x \in S$ remains in the cluster $C_x \subset D$ if there is a sample $y \in C_x$ such that}
\begin{equation} \label{filt_rule}
\mathcal{D}(y,c_x) + \tau \geq  \max \{\mathcal{D}(y,x),\mathcal{D}(x,c_x)\} 
\end{equation}

\noindent where $c_x$ is the centroid of the cluster $C_x$ and $\tau \geq 0$ is the parameter of our model. We used the Euclidean distance $\mathcal{D}$ throughout the experimental part. According to the decision rule \eqref{filt_rule}, $x$ will be added to the set $D$ clustered into \textit{known malware families}, i.e., families that appeared before the time $T$, or to a set clustered into \textit{new malware families}, which emerged after the time $T$. Samples from new malware families are clustered using an online clustering algorithm, such as OKM, SOM, or BSAS. Section \ref{online} presents the clustering results for these three algorithms. 

\begin{figure}
\centering
\includegraphics[scale=1.5]{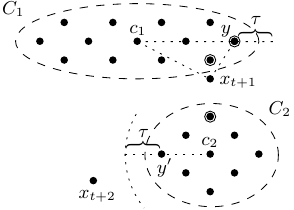}
\caption{Demonstration of the decision rule \eqref{filt_rule} used to determine whether the sample $x_{t+1}$ will remain in the nearest cluster $C_1 \subset D$ corresponding to a \textit{known malware family} and the sample $x_{t+2}$ will be assigned into cluster corresponding to a \textit{new malware family}. Three nearest neighbors of the sample $x_{t+1}$ are highlighted using the circle.}
\label{fig:2}
\end{figure}

The parameter $\tau > 0$ allows the clusters to expand. If $\tau < 0$, then we can extend to $C_x$ only by internal points, i.e., points closer to the centroid $c_x$ than the farthest point of the cluster $C_x$. Fig. \ref{fig:2} demonstrates the clustering decision rule \eqref{filt_rule} using the simple data set consisting of only two small clusters $C_1$ and $C_2.$ Fig. \ref{fig:2} shows that $x_{t+1}$ remains in $C_1$ since there is a sample $y \in C_1$ for which rule \eqref{filt_rule} is satisfied. On the other hand, $x_{t+2}$ will be clustered into a \textit{new malware family} (i.e., not belonging to the set $K$), because rule \eqref{filt_rule} is not satisfied even for $y'$, which is the best candidate for $y$.

The WKNN classifier was compared with the Multilayer perception and Random forest classifiers, and based on the classification results presented in Section 6.2, WKNN was selected to classify the streaming data in our proposed model. WKNN takes into account the similarities of the samples using mutual distance, and in addition to the KNN, the WKNN classifier considers the distances between nearest neighbors and the queried object. Note that a variant of the rule \eqref{filt_rule} was used in \cite{jurevcek2022parallel} to select a representative training set to train a classifier designed for malware detection.

The rest of this section presents the dataset used in the experimental part, and the metrics for evaluating clustering results. The implementation of our proposed model and methods for evaluating clustering results are based on scikit-learn \cite{scikitlearn} and PyClustering\footnote{\url{https://pyclustering .github.io}} libraries. All experiments in this work were executed on a single computer platform having two processors (Intel Xeon Gold 6136, 3.0GHz, 12 cores each), with 64 GB of RAM running the Ubuntu server 18.04 LTS operating system.

\subsection{Dataset}
\label{sec:dataset}

We evaluated our proposed method using EMBER dataset \cite{anderson2018ember}. 
The dataset contains 400,000 feature vectors corresponding to malicious samples from more than 3,000 malware families. The features were extracted using the LIEF open source package \cite{lief} and include metadata from portable executable file format \cite{microsoftPE}, strings, byte, and entropy histograms. The feature set consists of 2,381 features that are described in \cite{anderson2018ember}. These features were extracted using static analysis only, which aims at searching for information about the file structure without running a program. The dataset also contains feature vectors for benign samples which were not considered in our work. 

Date of the first appearence of the corresponding sample and the name of malware family where the sample belongs to are assigned to each feature vector. The date information is given by month and year of the first appearence of the sample. Samples that appeared until October 2018 are included in the EMBER training set, while samples appeared between November and December 2018 are included in the EMBER test set. While the EMBER training set contains samples from more than 3,000 malware families, we focus primarily on the four most prevalent malware families: Xtrat, Zbot, Ramnit, and Sality. The training dataset $D$ used in our model consists of samples from the EMBER training set with labels corresponding to these four malware families. The streaming data $S$ used in our model consists of samples from the EMBER test data set with labels corresponding to these four malware families and three additional malware families: Emotet, Ursnif, and Sivis. We considered three new families to get closer to the real situation when new malware families are constantly being created. One of our goals is to verify whether our proposed model can identify new families using online clustering.

Table \ref{table:dataset} summarizes the number of samples used in the experimental part, arranged in descending order of sample count for each of the seven prevalent malware families from the EMBER dataset. More information about malware families and technical details can be found in \cite{trend2023threat}.

\begin{table}[h] 
\centering
\begin{tabular}{ |l|c|c|c| }
\hline
Malware Family  & $|D|$ &  $|S|$& Size\\
\hline
Xtrat & 16,689 & 19,280 & 35,969\\
Zbot & 10,782 & 13,293 & 24,075\\
Ramnit & 10,275 & 10,320 & 20,595\\
Sality & 9,522 & 9,050 & 18,572\\
Ursnif & 0 & 5,733 & 5,733\\
Emotet & 0 & 4,904 & 4,904\\
Sivis & 0 & 2,803 & 2,803\\
\hline
\end{tabular}
\caption{The size of unlabeled data set $D$, size of streaming unlabeled data set $S$, and the overall data set size, i.e., $|D|+|S| = 47,268 + 65,383 = 112,651$.}
\label{table:dataset}
\end{table}

\subsection{Evaluation Metric}

We evaluated the quality of clusters using two standard measures: purity and silhouette coefficient (SC). Let the purity of cluster $C_j$ be defined as $\mathrm{Purity}(C_j) = \max_i p_{ij},$  where  $p_{ij}$ is the probability that a randomly selected sample from cluster $C_j$ belongs to class $i$. The overall purity is the weighted sum of individual purities and is given by
$$\mathrm{Purity} = \frac{1}{n}\sum_{j=1}^k{|C_j| \mathrm{Purity}(C_j)}.$$

\noindent where $n$ is the size of a dataset.

While purity uses labels when evaluating the quality of clusters, the silhouette coefficient does not depend on labels. It can therefore be used in the validation phase to determine the number of clusters.
The average silhouette coefficient \cite{rousseeuw1987silhouettes} for each cluster is defined as follows.

Consider $n$ samples $x_1,\ldots,x_n$ that have been divided into the $k$ clusters $C_1, \ldots, C_k.$ Average distance between $x_i \in C_j$ to all other samples in cluster $C_j$ is given by
$$
a(x_i) = \frac{1}{|C_j|-1}\sum_{\substack{y \in C_j \\ y \neq x_i}} \mathcal{D}(x_i,y).
$$

Let $b_k(x_i)$ be the average distance from the sample $x_i \in C_j$ to all samples in the cluster $C_k$ not containing  $x_i,$ and is defined by
$$
b_k(x_i) = \frac{1}{|C_k|}\sum_{y \in C_k} \mathcal{D}(x_i,y).
$$

Let $b(x_i)$ be the minimum of $b_k(x_i)$ for all clusters $C_k,$ where $k \neq j.$ The silhouette coefficient of $x_i$ is given by combining $a(x_i)$ and $b(x_i)$, and is defined by
$$
s(x_i) = \frac{b(x_i)-a(x_i)}{\max(a(x_i),b(x_i))} .
$$

The silhouette coefficient $s(x_i)$ ranges from -1 to 1, with higher scores indicating better performance. Finally, the average silhouette coefficient for a given dataset is defined as the average value of $s(x_i)$ over all samples in the dataset.

Note that our proposed model assumes that only unlabeled data is available. Therefore, for example, when choosing the optimal number of features, the silhouette coefficient is used, while purity is used only to evaluate our model.

\section{Experimental Results} \label{evaluation}

This section contains descriptions of experiments conducted for our proposed online clustering model. Firstly, we experiment with the number of features used to represent the samples from malware families. Then, we select a machine learning algorithm to classify streaming data to the known malware families and tune the parameter $\lambda$ of our proposed model, enabling cluster expansion. Finally, we present the experimental results of our proposed online clustering model, compare it with the reference model, demonstrate that the computational times are low enough to cluster all malware samples that appear daily, and provide a discussion for our work.

\subsection{Preprocessing and Clustering Algorithm Selection} \label{clust}

The preprocessing used in this work consists of data normalization and dimensionality reduction. Dataset $D$ was normalized using the standard score, and the PCA algorithm was used to extract optimal features from the original features. When calculating the standard score of the streaming data $S = \{x_{t +1}, x_{t+2}, \ldots\}$, standard deviations and mean values were obtained based on dataset $D$. Then, the PCA transformation was applied to the normalized data, where the PCA transformation was created based on dataset $D$.

In this experiment, we considered options for the optimal number of features from the set $\{20, 30, 40, \ldots, 80\}$. We experimented with the following three clustering algorithms for clustering the dataset $D$: $k$-means, SOM, and DBSCAN. The optimal number of features was chosen via the silhouette coefficient which was used to evaluate the clusters created by the three clustering algorithm. The number of cluster was set to four since the dataset $D$ consists of samples from four malware families. 

The number of clusters determined the number of output neurons in SOM. We left all other SOM hyperparameters at their default values according to the PyClustering library. The implementation of the DBSCAN and the $k$-means is based on the scikit-learn library. We tuned two hyper-parameters of the DBSCAN using the following search grid:
\begin{itemize}
\item eps: 0.1, 0.5, 1, 2, 5
\item min\_samples: 5, 10, 20
\end{itemize}

The parameter eps is defined as the maximal distance between samples, where one is considered to be in the neighborhood of the other one. The parameter min\_samples is the minimum number of points that are required to form a dense region. The highest silhouette coefficient for the DBSCAN was achieved for eps $= 5$, and min\_samples $= 10$.

\begin{figure}[htbp]
\centering
\includegraphics[scale=0.5]{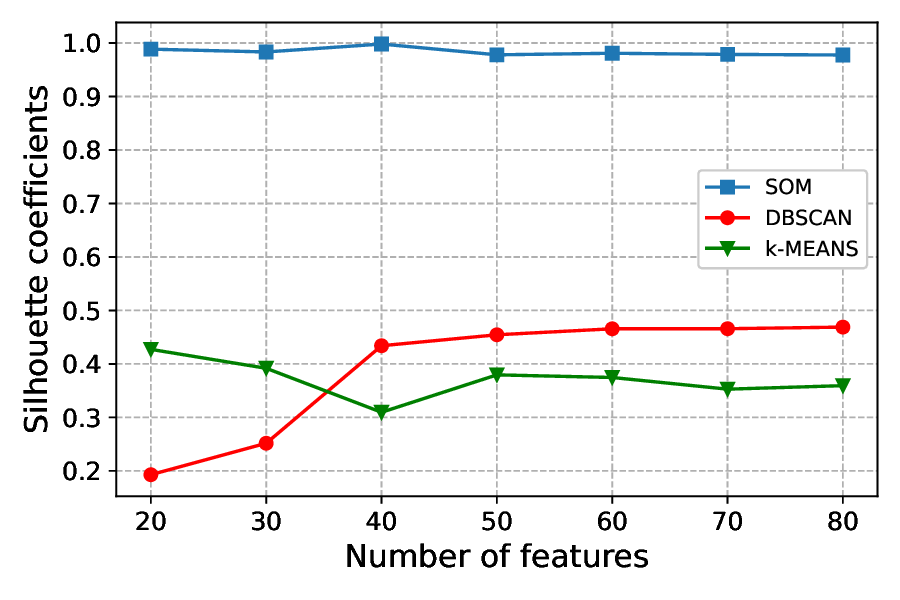}
\caption{The relationship between the number of features and the silhouette coefficient.}
\label{fig:3}
\end{figure}

Fig. \ref{fig:3} shows the relation between the number of features extracted by PCA and the average silhouette coefficient for three clustering algorithms. The highest silhouette coefficient was achieved for 40 extracted features by SOM. Note that the highest purity of clusters, 84.46\%,  was achieved for 50 features by DBSCAN. Since we assumed that the dataset contains only unlabeled samples, we used 40 features for all remaining experiments from this work.

\subsection{Classifier Selection and Tuning of the Hyper-parameter $\tau$}

To classify streaming data to the known malware families, we considered the following three classifiers: Multilayer perceptron (MLP),
Random forest (RF), and $k$-nearest neighbors (KNN). MLP \cite{lecun2015deep} is an artificial neural network composed of multiple layers of neurons, typically including an input layer, one or more hidden layers, and an output layer. The input layer takes an input, which is then processed in hidden layers, and finally, perceptrons in the output layer output a result. Random forest \cite{breiman2001random} is an ensemble learning method combining the results made by several decision trees using a voting mechanism. The $k$-nearest neighbors classifier \cite{cover1967nearest} is a non-parametric method that predicts a class label according to a majority vote of its $k$ nearest neighbors. 

We tuned the hyper-parameters of the MLP, RF, and KNN
classifiers using the grid search that exhaustively considered all parameter combinations. The following searching grid parameters were explored for MLP: 
\begin{itemize}
	\item hidden layer sizes: (100,0), (200,0), (400,0), (100,50), (200,100), (400,100), (400,200)
	\item activation function: relu, tanh, logistic
	\item solver for weight optimization: lbfgs, adam
	\item alpha: 0.0001, 0.001, 0.01
\end{itemize}
The parameter alpha controls the strength of regularization applied to the neural network's weights. The definitions of the activation functions and the solvers are presented in the neural\_network.MLPclassifier class from the scikit-learn library, which was used in our experiments. For random forest, we explored the number of trees in the forest, the maximal depth of trees, and the criterion that measure the quality of a split: 
\begin{itemize}
	\item number of estimators: 100, 500, 1000
	\item maximal depth: 7, 8, 9, 10
	\item criterion: gini, entropy
\end{itemize}

The criteria are defined in the ensemble.RandomForestClassifier class from the scikit-learn library, which was used in our experiments. Finally, for the KNN, we considered the following hyper-parameters:
\begin{itemize}
	\item $k$: 1,3,5,7,9,11
	\item weights: uniform, wknn
\end{itemize}

\begin{table*}
\setlength{\tabcolsep}{9pt} 
\begin{center}

\resizebox{\textwidth}{!}{

\begin{tabular}{|l|c|c|c|c|c|c|c|c|c|} 
\hline 
classifiers & \multicolumn{4}{|c|}{ MLP} &
 \multicolumn{3}{|c|}{ RF } & \multicolumn{2}{|c|}{ KNN }\\
\hline
\multicolumn{1}{|l|}{parameters} &  hidden\_layer\_sizes & activation & solver & alpha &  criterion &   max\_depth & n\_estimators &   $k$ &   weights  \\
\hline
\multicolumn{1}{|l|}{best-performing values}  &  (400, 200)  & relu & adam  & 0.0001  &    entropy & 10 & 1000  & 3 & wknn\\    
\hline
classification accuracies & \multicolumn{4}{|c|}{ 93.89\%} &
 \multicolumn{3}{|c|}{ 92.31\% } & \multicolumn{2}{|c|}{ 94.08\% }\\
\hline
\end{tabular}

}
\caption{Hyperparameter tuning for the MLP, RF, and KNN classifiers.} \label{tab:tuning}

\end{center}
\end{table*}  

The parameter $k$ denotes the numbers of nearest neighbors, and the parameter weights denotes the weight function. Uniform weight states that all $k$ neighbors are weighted equally, while the case "weights=wknn" is described in \eqref{wknn}. The best-performing values of the hyperparameters for the MLP, RF, and KNN models, together with the corresponding classification accuracies, are given in Table \ref{tab:tuning}. Since the WKNN classifier achieved the highest classification accuracy, we used it in all remaining experiments.

The proposed online clustering model has the parameter $\tau$ enabling cluster expansion. We experimented with the $\tau$ values from the set $\{-5, -2, 0, 2, 5\}$ to determine the optimal values. The highest silhouette coefficient was achieved for $\tau = -2$ where 11.2\% of samples from $S$ were determined according to the clustering decision rule \eqref{filt_rule} as samples from \textit{new malware families}. 

\begin{figure}[htbp]
\centering
\includegraphics[scale=0.5]{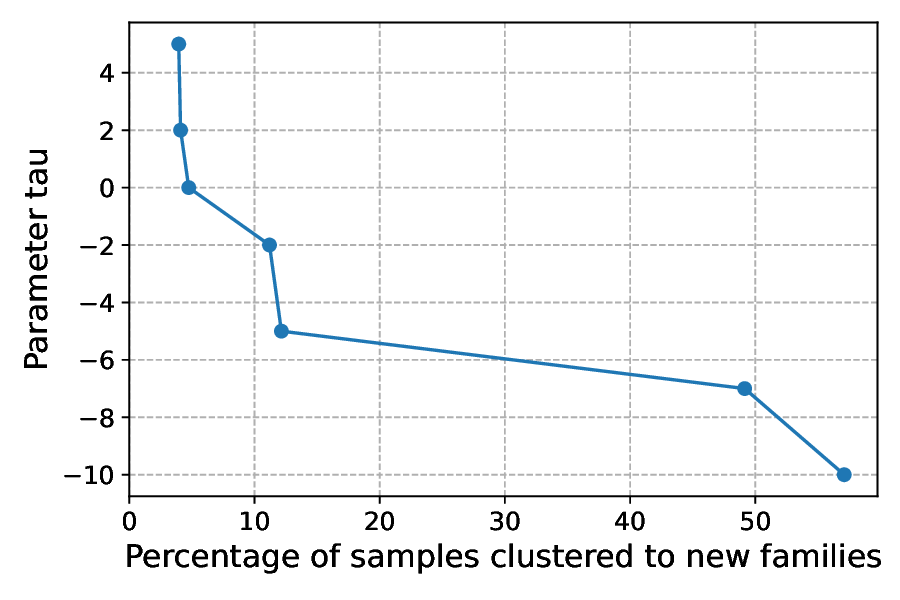}
\caption{The relationship between the parameter $\tau$ and the percentage of streaming data clustered to \textit{new malware families}.}
\label{fig:4}
\end{figure}

Based on the rule \eqref{filt_rule}, with increasing $\tau$, the number of elements clustered into \textit{known malware families} increases. Fig. \ref{fig:4} shows how the parameter $\tau$ influences the number of samples from the streaming dataset classified as samples from \textit{new malware families}. The figure was created for the parameter $k=3$ of the WKNN classifier, and we assumed that the number of clusters in dataset $D$ equals four.

\subsection{Online Clustering} \label{online}

This section describes the experimental results of the proposed online clustering model. We evaluated the model using three state-of-the-art online clustering algorithms: SOM, BSAS, and OKM. We applied these algorithms to cluster samples determined by the clustering decision rule \eqref{filt_rule} as samples belonging to \textit{new malware families}. Since the number of newly emerging malware families during a specific time window is unknown, we assume that the correct number of clusters, which is required information for SOM and OKM clustering algorithms,  is also unknown. While the number of clusters is not required in BSAS, the upper bound for the number of clusters must be provided. The following experiments were conducted for the number of clusters (or its upper bound for BSAS) in the set $\{4, 5, \ldots, 10\}$. 

\begin{figure*}
\begin{subfigure}[b]{0.47\linewidth}
\centering
\includegraphics[ width=\linewidth]{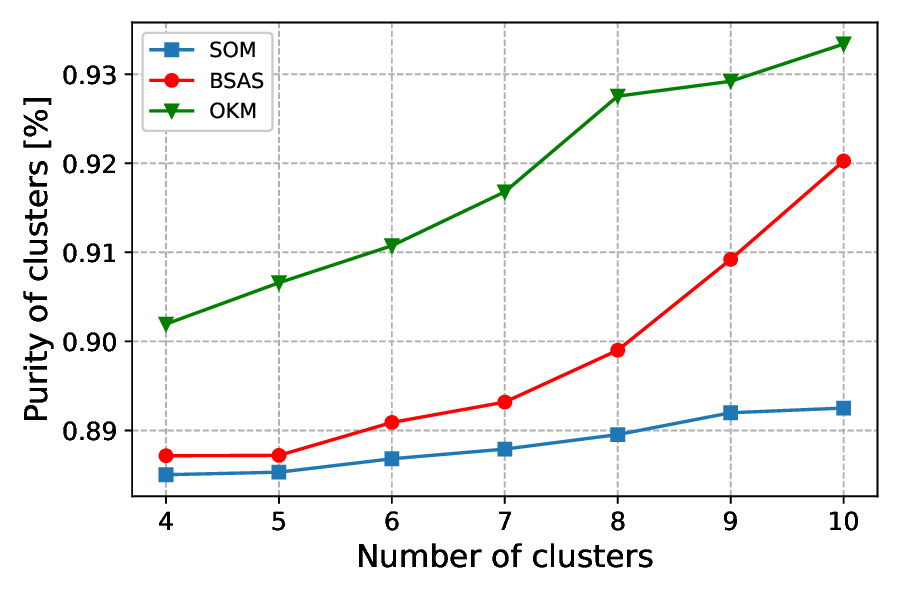}
\caption{Purities of clusters.}
\end{subfigure}
\hfill
\begin{subfigure}[b]{0.47\linewidth}
\centering
\includegraphics[width=\linewidth]{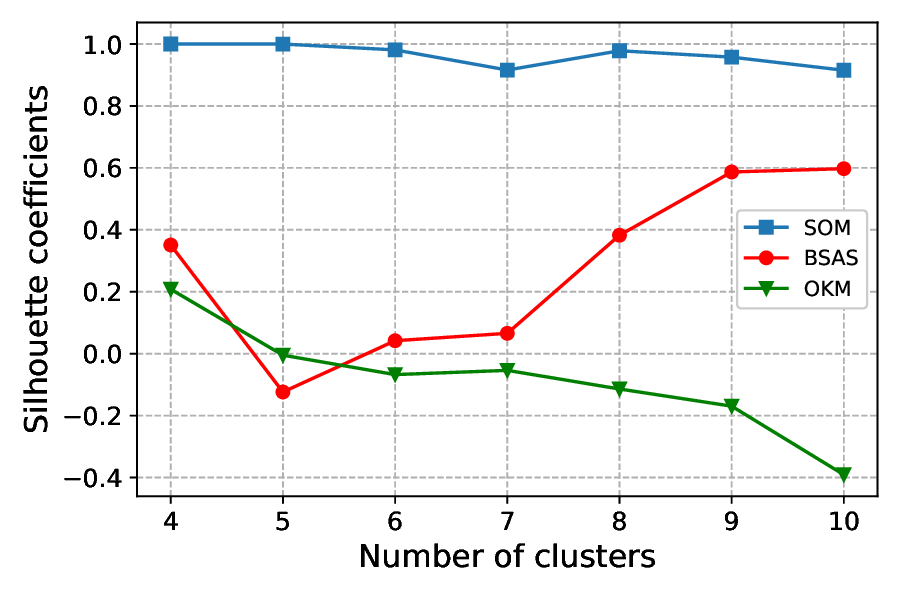}
\caption{Average silhouette coefficients.}
\end{subfigure}
\caption{The relation between the number of clusters and the purity of clusters (a), respectively, the average silhouette coefficient (b). The results correspond to samples that were clustered to \textit{new malware families}.} \label{fig:t06}
\end{figure*}

We applied SOM, BSAS, and OKM twenty times to samples clustered into \textit{new malware families}, and Fig. \ref{fig:t06} shows the average results for the purity of clusters and the silhouette coefficient, considering various numbers of clusters. These clustering results correspond to the parameter $\tau = -2$, for which our model achieved the highest silhouette coefficient on the dataset $D$. 

The clustering results show that all three online clustering algorithms achieved a purity of clusters of at least 88.5\%, with OKM outperforming both BSAS and SOM. However, SOM achieved a significantly higher average silhouette coefficient than BSAS and OKM. The average silhouette coefficient values close to 1 indicate that the clusters are well-separated. The clustering results show that the highest purity of clusters, 93.34\%, was achieved using OKM for ten clusters, and the highest average silhouette coefficient, 0.99, was performed using SOM for four clusters. In the previous work \cite{jureckova2023classification},  the SOM also achieved significant results compared to  BSAS and OKM online clustering algorithms.

\begin{figure*}
\begin{subfigure}[b]{0.47\linewidth}
\centering
\includegraphics[ width=\linewidth]{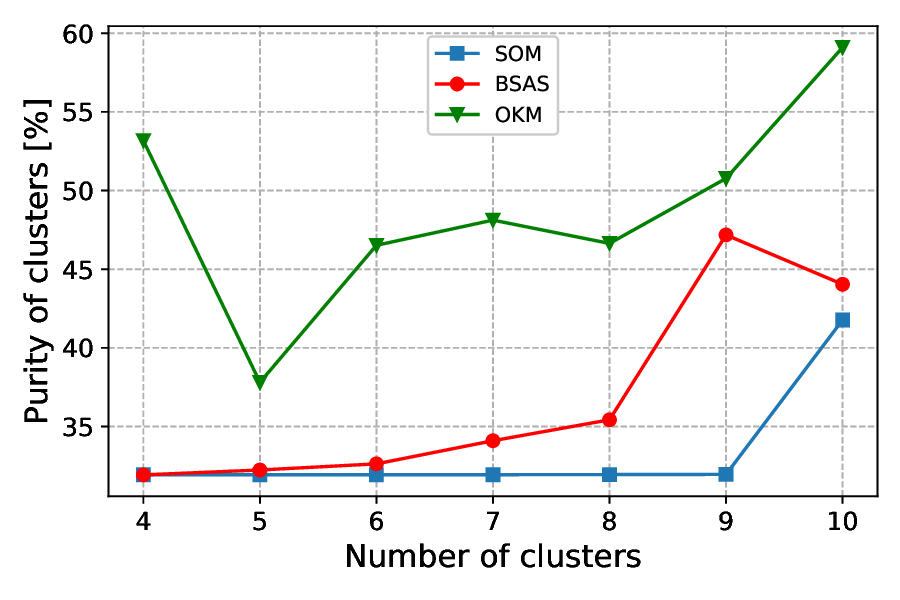}
\caption{Purities of clusters.}
\end{subfigure}
\hfill
\begin{subfigure}[b]{0.47\linewidth}
\centering
\includegraphics[width=\linewidth]{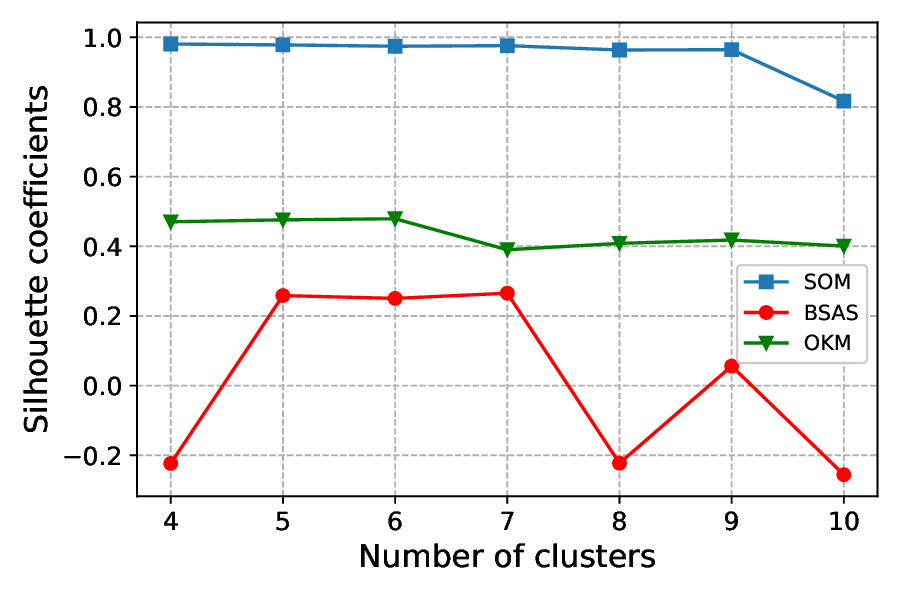}
\caption{Average silhouette coefficients.}
\end{subfigure}
\caption{The relation between the number of clusters and the purity of clusters (a), respectively, the average silhouette coefficient (b). The online clustering algorithms were directly applied to the unlabeled dataset $D$ and the streaming data $S$.} \label{fig:t07}
\end{figure*}

The average silhouette coefficient and purity of clusters calculated for samples from \textit{known families} are 0.99 and 56.59\%, respectively, where the purity is significantly lower using the parameters $\tau = -2$ in comparison to this metric calculated for samples from new malware families. 

Finally, we compare the proposed online clustering model with the reference model, where the online clustering algorithms were directly applied to the unlabeled dataset $D$, and the streaming data $S$ consisting of a total of 112,651 malicious samples from seven prevalent malware families. Fig. \ref{fig:t07} shows the purities of clusters and the average silhouette coefficient achieved for the unlabeled dataset $D$ and the streaming data $S$ for several numbers of clusters. The results indicate that the proposed online clustering model is more successful in terms of purity of clusters than the approach where we directly apply the online clustering algorithm.

\subsection{Computational Times}

\begin{figure*}
\begin{subfigure}[b]{0.3\linewidth}
\centering
\includegraphics[ width=\linewidth]{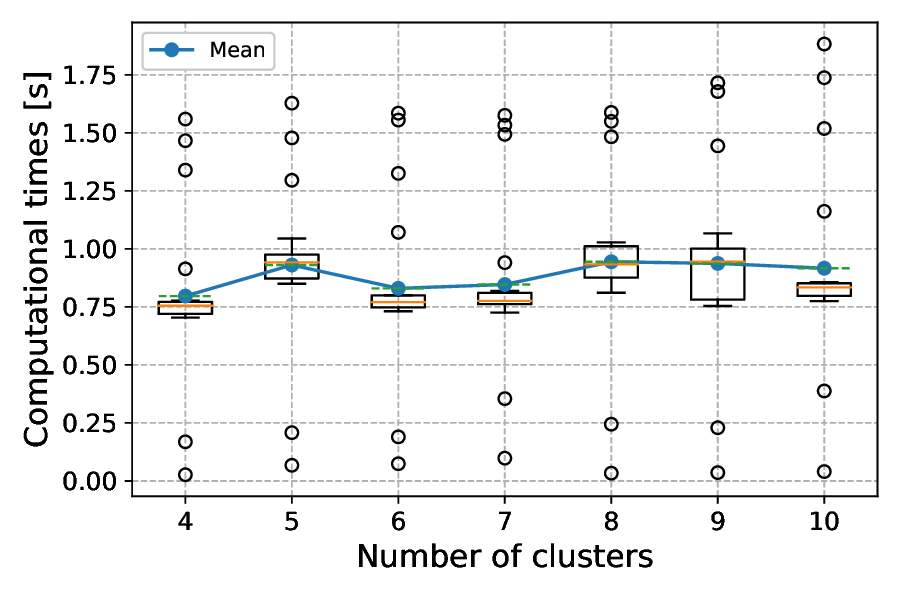}
\caption{SOM}
\end{subfigure}
\hfill
\begin{subfigure}[b]{0.3\linewidth}
\centering
\includegraphics[width=\linewidth]{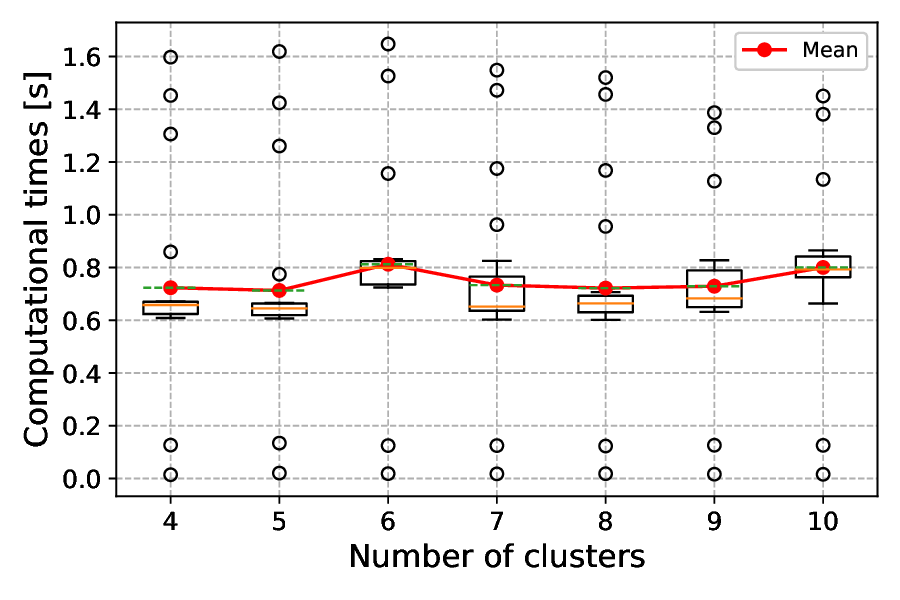}
\caption{BSAS}
\end{subfigure}
\hfill
\begin{subfigure}[b]{0.3\linewidth}
\centering
\includegraphics[ width=\linewidth]{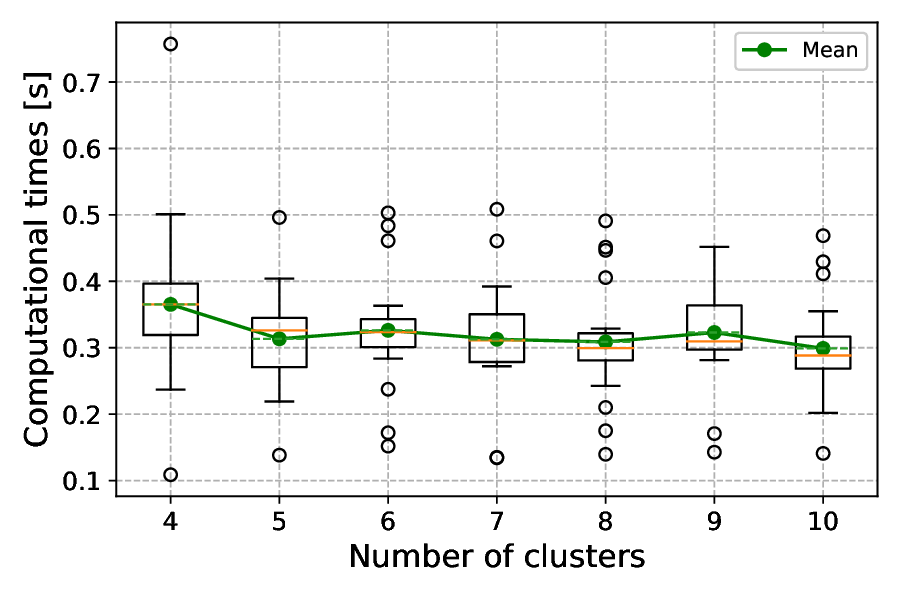}
\caption{OKM}
\end{subfigure}
\caption{The computational times of the online clustering algorithms.} \label{fig:times_clustmax}
\end{figure*}

This section presents the computational times of three online clustering algorithms, SOM, BSAS, and OKM, applied to cluster the samples determined by the clustering decision rule \eqref{filt_rule} as samples belonging to \textit{new malware families}. We run these algorithms twenty times, presenting the results as boxplot graphs. The average number of samples clustered using the online clustering algorithms is 3,505, with a standard deviation 776. This number of samples is based on the WKNN classification results and the clustering decision rule as described in Section \ref{proposed}. Fig. \ref{fig:times_clustmax} shows the computational times of individual online clustering algorithms. The mean values of computational times for clustering all samples belonging to \text{new malware families} are less than one second for all clustering algorithms and all considered numbers of clusters. The graphs also demonstrate that the OKM algorithm is the fastest among the three online clustering algorithms, whereby SOM is approximately two times slower than OKM.

\begin{figure}[htbp]
\centering
\includegraphics[scale=0.5]{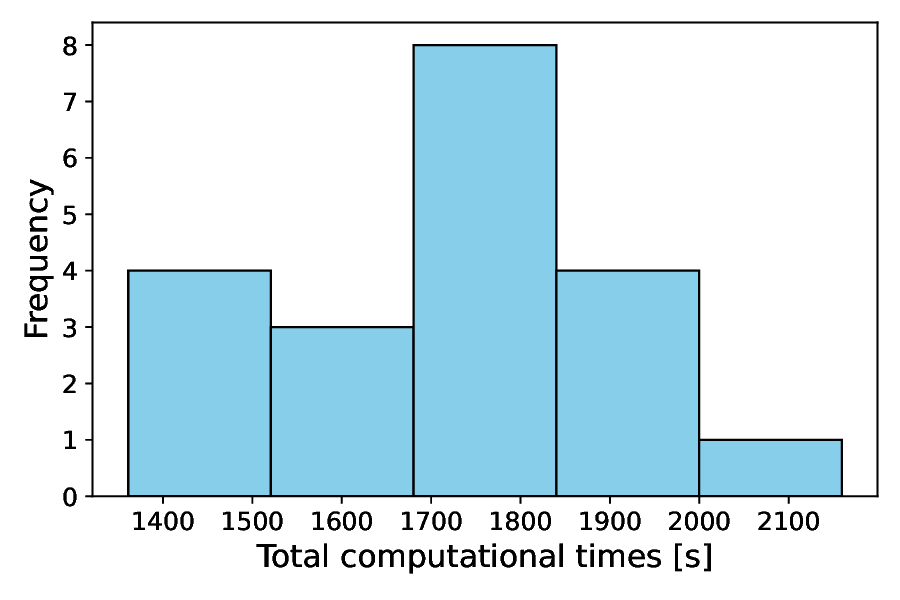}
\caption{Histogram of average computational times for 20 measurements of the entire proposed model.}
\label{pic:hist}
\end{figure}

It took less than 1 second to cluster the unlabeled data $D$ , i.e., the data that appeared before the streaming data $S$. The training of the WKNN classifier took 1,706 seconds on average, with a standard deviation of 206 seconds. The total computational time of the proposed model consists of computational times for data preprocessing, clustering of the dataset $D$, WKNN classification of the streaming data $S$, and online clustering of samples belonging to \textit{new malware families}, and is shown in Fig. \ref{pic:hist} in the form of a histogram. 

Note that the clustering of the dataset $D$ is conducted only once. Unlike other classifiers, such as neural networks or support vector machines, WKNN does not learn a discriminative function from the training data. As a result, training of WKNN is done sequentially as streaming data comes. If we used, for example, a neural network to classify malware families, the training would be performed only once, which could reduce the total computation time of the proposed model. On the contrary, the advantage of the WKNN classifier is that it does not need to be retrained; however, distances between testing and training samples must be computed, which might be computationally expensive for large datasets.

According to the AV-Test Institute \cite{avtest2024avtest}, 450,000 new malware samples are detected on average daily. Based on the computation times shown in Fig. \ref{pic:hist}, all malware samples that appear daily can be clustered using the proposed model for online stream data processing. Specifically, the mean of computational times from Fig. \ref{pic:hist} is 1,728 seconds, the average computational time for clustering the streaming data $S$ consisting of 65,383 samples. As a result, processing 450,000 samples would take approximately 3.3 hours. For the highest computational time, 2,159 seconds, for processing the streaming data, processing the 450,000 samples would take approximately 4.13 hours.

\subsection{Discusion}

This work deals with the problem of online clustering of streaming data concerning a fixed dataset consisting of unlabeled samples that appeared before the streaming data (the problem is defined in the first paragraph of Section \ref{proposed}). This problem differs from the straightforward application of clustering algorithms to a single fixed or streaming data in that we also use another dataset consisting of older samples to improve clustering results. This aligns with the real situation when antivirus companies have to analyze streaming data while also having older, unlabeled data. If the older dataset contains a subset of labeled samples, in this case, we could use semi-supervised learning techniques, with the help of which we could improve the online clustering of streaming data.

The proposed model for the online clustering works with malware samples only. If the streaming data contains benign and malicious samples, applying a malware detection model before clustering into malware families using our model will be necessary. 

\section{Conclusion} \label{conclusion}

Clustering malware samples into families is suitable for speeding up the work of malware analysts and also for research purposes. Clustering malware families allows us to examine the evolution of individual malware families over time and potentially help with the prediction of future variants of malware. In this work, we proposed a model for the online clustering of malicious samples into malware families. Streaming data is not clustered directly but split according to similarity with samples from known malware families. The samples that the proposed system determined did not belong to existing families were clustered into emerging families using online clustering algorithms. The clustering results show that the online clustering algorithms achieved a purity of clusters of at least 88.5\%. Experimental results indicate that this approach creates clusters with higher purity than clusters formed by the direct application of an online clustering algorithm.

Future work may focus on testing and possibly improving the proposed model to cluster incoming samples from a more significant number of families with the required purity of clusters. This task is challenging since the feature set obtained from the static analysis may not be sufficient to distinguish a more significant number of families from each other. Another extension of the work can be using semi-supervised learning methods, which could improve clustering into malware families utilizing a subset of labeled samples.

\section*{Acknowledgements}
This work was supported by the OP VVV MEYS funded project CZ.02.1.01/0.0/0.0/16\_019/0000765 "Research Center for Informatics" and by the Grant Agency of the CTU in Prague, grant No. SGS23/211/OHK3/3T/18 funded by the MEYS of the Czech Republic.

\bibliography{template}

\end{document}